\documentclass{article}
\usepackage{spconf,amsmath,amssymb,graphicx,pifont,booktabs,cite}
\usepackage[hidelinks]{hyperref}
\usepackage{fancyhdr}
\fancypagestyle{arxiv}{%
  \fancyhf{}%
  \fancyfoot[C]{\parbox{\textwidth}{%
    \fontsize{9}{10.5}\selectfont\raggedright
    This work has been submitted to the IEEE for possible publication. Copyright may be transferred without notice, after which this version may no longer be accessible.}}%
}
\newcommand{\cmark}{\ding{51}}
\newcommand{\xmark}{\ding{55}}
\let\oldthebibliography\thebibliography
\renewcommand{\thebibliography}[1]{\oldthebibliography{#1}%
  \setlength{\itemsep}{0pt plus 0.3ex}\setlength{\parsep}{0pt}}
\title{Q-SPT: Learnable Query-Based Compression for Low-Frame-Rate Speech Tokenization}
\name{Jeeyoung Yun \qquad Seohwan Yun \qquad Sungwoong Kim$^{\dagger}$%
  \thanks{\small $^{\dagger}$Corresponding author.}}
\address{Korea University, Seoul, Republic of Korea \\
  \texttt{\{jee010910, jitoo6342, swkim01\}@korea.ac.kr}}
\begin{document}
\ninept
\maketitle
\thispagestyle{arxiv}
\begin{abstract}
Neural speech codecs increasingly serve as tokenizers for speech language models (SLMs).
Lowering the frame rate reduces the computational and memory costs of SLMs, but makes it difficult to preserve both linguistic information and acoustic detail.
Existing approaches rely on rule-based compression: average pooling can discard linguistic information, whereas similarity-based merging uses a fixed threshold on adjacent-frame similarity and applies the resulting boundaries to the acoustic stream.
We propose Q-SPT, a low-frame-rate dual-stream speech tokenizer with separate, context-aware, learnable query-based compressors specialized for semantic and acoustic representations.
In particular, queries at a fixed rate independently attend to the semantic and acoustic streams as separate key–value sources, enabling stream-specific, context-aware aggregation through two separately learned compressors.
In addition, an autoregressive text loss explicitly supervises the semantic compressor to preserve linguistic information.
Experimental results show that Q-SPT achieves the best reconstruction among the evaluated codecs at the same frame rate. In downstream SLMs, it yields the best speech recognition accuracy and text-to-speech perceptual quality with competitive intelligibility.
\end{abstract}


\section{Introduction}
\label{sec:introduction}
Neural speech codecs compress speech into compact discrete tokens and reconstruct waveforms while preserving perceptual quality~\cite{encodec, dac, soundstream}. These codecs increasingly serve as tokenizers for speech language models (SLMs), enabling speech understanding and generation through language modeling over discrete speech tokens~\cite{moshi, valle}. Effective speech tokenization therefore requires representations that support both high-quality waveform reconstruction and downstream speech modeling.

Recent codecs represent speech with fewer tokens by reducing the number of codebooks per frame~\cite{hificodec, wavtokenizer} or the temporal resolution~\cite{moshi, lfsc}, while maintaining reconstruction quality. The reduced token rate also improves the efficiency of SLMs by lowering the computational and memory costs of both training and inference~\cite{lfsc, moshi}. However, at extremely low frame rates, each token must carry more speech content, making it difficult to preserve both linguistic information and acoustic detail~\cite{flexicodec}.

To enhance linguistic information while retaining acoustic detail, prior codecs leverage pretrained speech representations. SpeechTokenizer~\cite{speechtokenizer} and Mimi~\cite{moshi} distill self-supervised learning (SSL) features into the first residual vector quantization (RVQ)~\cite{soundstream} layer, while X-Codec~\cite{xcodec} fuses semantic and acoustic features before quantization. More recently, dual-stream codecs encode semantic and acoustic features in separate streams to preserve both~\cite{dualcodec, flexicodec}. DualCodec~\cite{dualcodec} downsamples SSL features to a fixed frame rate by average pooling, but this uniform averaging can discard linguistic information at extremely low frame rates, degrading intelligibility~\cite{lmspt, flexicodec}. 
FlexiCodec~\cite{flexicodec} instead merges adjacent frames based on semantic similarity, improving information preservation at average frame rates as low as 6.25\,Hz. However, the same merging boundaries are applied to the acoustic stream, although semantically similar frames may still differ in acoustic detail.
Moreover, the merging rule relies only on adjacent-frame similarity with a fixed threshold, so it neither uses broader temporal context nor is optimized by the training objective.

\begin{figure*}[t]
    \centering
    \includegraphics[width=0.9\textwidth]{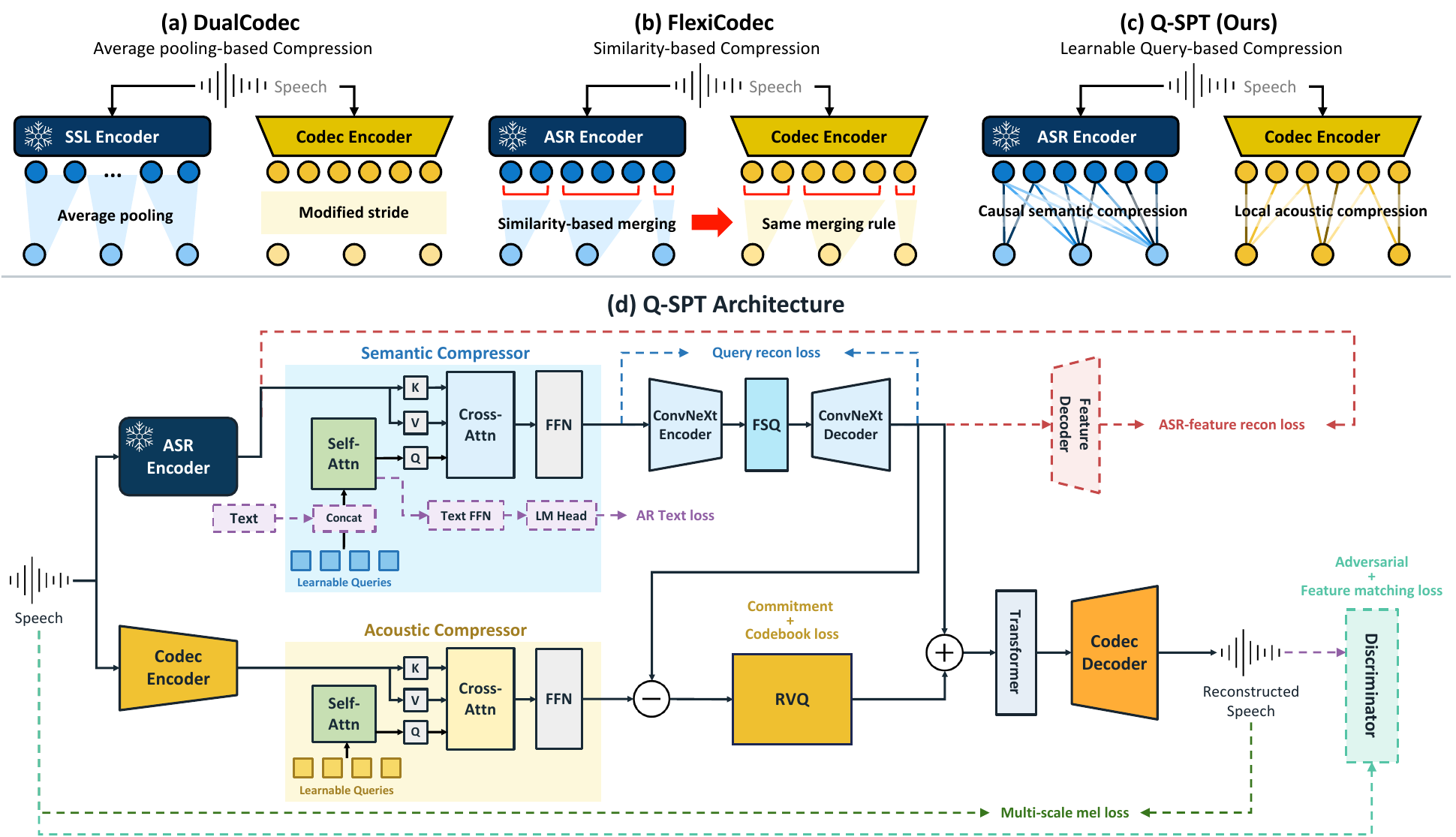}
    \caption{Comparison of temporal compression strategies for low-frame-rate speech tokenization:
    (a) average pooling in DualCodec~\cite{dualcodec},
    (b) similarity-based merging in FlexiCodec~\cite{flexicodec},
    and (c) learnable query-based compression in Q-SPT.
    (d) Detailed architecture and training objectives of Q-SPT.
    Dashed elements indicate components and connections used only during training.}
    \label{fig:fig1}
\end{figure*}

We propose \textbf{Q-SPT}, a low-frame-rate dual-stream speech tokenizer based on learnable query-based compression. Unlike rule-based compression such as average pooling or similarity-based merging (Fig.~\ref{fig:fig1}(a)--(b)), Q-SPT compresses features through cross-attention with learnable queries, which adaptively weight frames within each temporal window (Fig.~\ref{fig:fig1}(c)).
Since the semantic and acoustic streams carry different information, Q-SPT employs separate semantic and acoustic compressors, each with a distinct temporal context accessible to its fixed-rate queries.
In addition, because reconstruction objectives supervise linguistic content only implicitly, we add an autoregressive text loss that explicitly supervises the semantic compressor.
By conditioning autoregressive transcript prediction on the semantic queries, this loss encourages the semantic tokens to retain linguistic information for downstream SLMs.

Our contributions are summarized as follows:
\begin{itemize}
    \item We present Q-SPT, which replaces rule-based temporal compression in dual-stream speech tokenization with \textbf{learnable query-based compression specialized for each stream}.
    Separate semantic and acoustic compressors with distinct temporal contexts adaptively weight frames to preserve both linguistic information and acoustic detail.
    \item We introduce an \textbf{autoregressive text loss} that explicitly supervises the semantic compressor through transcript prediction, which improves the preservation of linguistic information in the semantic tokens and leads to gains in both reconstruction and downstream SLM performance.
    \item Experiments show that Q-SPT achieves \textbf{the best reconstruction} among the evaluated codecs at the same frame rate and, in downstream SLMs, the best automatic speech recognition (ASR) performance and text-to-speech (TTS) perceptual quality.
\end{itemize}

\section{Method}
\label{sec:method}
Fig.~\ref{fig:fig1}(d) illustrates the architecture of Q-SPT. A frozen ASR encoder~\cite{sensevoice} and a DAC-based codec encoder~\cite{dac} extract semantic and acoustic features from speech, respectively. Separate semantic and acoustic compressors use learnable queries to compress the two feature sequences to 6.25\,Hz. Semantic tokens are obtained by finite scalar quantization (FSQ)~\cite{fsq}, and acoustic tokens are obtained by applying RVQ to the residual between the acoustic and quantized semantic representations~\cite{dualcodec}. The sum of the two quantized representations is upsampled by a Transformer with mask tokens, and the DAC-based codec decoder reconstructs speech from the upsampled representation.
During training, an autoregressive text loss additionally supervises the semantic compressor through transcript prediction.

\subsection{Learnable Query-Based Compression}
\label{sec:compression}
Given a 16-kHz waveform $\mathbf{x}\in\mathbb{R}^{L}$, a frozen SenseVoice-Small~\cite{sensevoice} ASR encoder extracts semantic features $\mathbf{F}^{s}\in\mathbb{R}^{T_s\times d}$ at approximately 16.7\,Hz, and the codec encoder extracts acoustic features $\mathbf{F}^{a}\in\mathbb{R}^{T_a\times d}$ at 12.5\,Hz. Here, $T_s$ and $T_a$ denote the sequence lengths of the semantic and acoustic features, respectively, and $d$ is the feature dimension. Each compressor is a stack of Transformer layers in which learnable queries aggregate its input features within a sliding window through cross-attention~\cite{blip2}. Both compressors output $\mathbf{H}^{s},\mathbf{H}^{a}\in\mathbb{R}^{T\times d}$ at a common frame rate of 6.25\,Hz, where $T=\lceil T_a/2\rceil$ is the compressed sequence length.

\noindent\textbf{Causal semantic compression.}
The semantic compressor uses a wide window with causal attention to capture linguistic information over a broader temporal context. 
Specifically, the $i$-th query ($i=1,\ldots,T$) cross-attends to a window of $w_s$ semantic frames ending at the $e_i$-th frame, where $e_i=\lceil iT_s/T\rceil$ is the last semantic frame within its time span. This window moves with the query index and is truncated at the utterance start, forming a causal sliding window. We set $w_s=16$, which corresponds to approximately 0.96\,s. 
Causal self-attention over the query sequence additionally provides preceding context beyond the local cross-attention window. To preserve the relative temporal order of queries and semantic frames, we apply rotary position embeddings (RoPE)~\cite{rope} to both attentions.

\noindent\textbf{Local acoustic compression.}
The acoustic compressor uses a narrow window to retain fine-grained acoustic detail.
Specifically, the $i$-th query cross-attends to a window of $w_a$ acoustic frames starting at the $(2i-1)$-th frame. We set $w_a=3$, which corresponds to approximately 0.24\,s.

\noindent\textbf{Quantization and waveform reconstruction.}
The semantic representation $\mathbf{H}^{s}$ is quantized into $\widehat{\mathbf{H}}^{s}\in\mathbb{R}^{T\times d}$ by FSQ~\cite{fsq} with temporal ConvNeXt layers~\cite{convnext}. 
Following DualCodec~\cite{dualcodec}, seven RVQ stages quantize the residual $\mathbf{H}^{a}-\widehat{\mathbf{H}}^{s}$ into $\widehat{\mathbf{H}}^{a}\in\mathbb{R}^{T\times d}$.
For reconstruction, $\widehat{\mathbf{H}}^{s}+\widehat{\mathbf{H}}^{a}$ is upsampled from 6.25 to 12.5\,Hz by interleaving learnable mask tokens and refining the sequence with a one-layer Transformer. The DAC-based codec decoder~\cite{dac} then reconstructs the waveform $\widehat{\mathbf{x}}\in\mathbb{R}^{L}$ from the upsampled features.

\subsection{Training with Autoregressive Text Supervision}
\label{sec:objective}
Q-SPT is trained end-to-end with the ASR encoder kept frozen. In addition to the standard codec losses for waveform reconstruction~\cite{dac, dualcodec}, we use two types of losses to enhance linguistic information in the semantic stream. An autoregressive text loss provides explicit supervision on linguistic content, whereas semantic reconstruction losses implicitly guide the semantic tokens to retain the information in the ASR features.

\noindent\textbf{Autoregressive text loss.}
Since reconstruction objectives provide only implicit supervision for linguistic content, we explicitly supervise the semantic compressor with the ground-truth transcript of the input speech $\mathbf{x}$. During training, the transcript is tokenized into $\mathbf{y}=(y_0,y_1,\ldots,y_M)$ with the BERT tokenizer~\cite{bert}, where $y_0$ is a start token and $M$ is the number of target tokens. The embedded tokens are concatenated after the semantic queries, and the resulting sequence is processed by the semantic compressor. Queries and text tokens share the same self-attention layers, whereas text tokens bypass cross-attention and use a separate text feed-forward network (FFN). Queries attend causally only among themselves and cannot attend to text tokens. In contrast, each text token attends to all queries and causally to the text tokens, so the loss supervises the compressor without leaking text into the query states. The text FFN outputs are then projected by an LM head to predict each next token. The autoregressive text loss is the next-token cross-entropy,
\begin{equation}
    \mathcal{L}_{\mathrm{AR}}
    = -\frac{1}{M}\sum_{m=1}^{M}\log p_{\theta}\!\left(y_m \mid \mathbf{x}, \mathbf{y}_{<m}\right),
\label{eq:ar_loss}
\end{equation}
where $p_{\theta}$ denotes the predicted distribution from the LM head, conditioned on $\mathbf{x}$ through the semantic queries and on the preceding tokens $\mathbf{y}_{<m}$.

\noindent\textbf{Semantic reconstruction losses.}
To further enhance linguistic information in the semantic tokens, we use two L2 reconstruction losses. The query reconstruction loss $\mathcal{L}_{\mathrm{QRec}}$ matches the quantized output $\widehat{\mathbf{H}}^{s}$ to the pre-quantization representation $\mathbf{H}^{s}$, so that the information shaped by the text loss is retained through FSQ. It is defined as
\begin{equation}
    \mathcal{L}_{\mathrm{QRec}}
    = \left\|\widehat{\mathbf{H}}^{s}-\mathrm{sg}(\mathbf{H}^{s})\right\|_2^2,
\label{eq:qrec}
\end{equation}
where $\mathrm{sg}(\cdot)$ denotes the stop-gradient operator. The ASR-feature reconstruction loss $\mathcal{L}_{\mathrm{FRec}}$ acts similarly to semantic distillation~\cite{speechtokenizer, dualcodec}. A learnable convolutional feature decoder $D$ expands $\widehat{\mathbf{H}}^{s}$ along the time axis to the length of $\mathbf{F}^{s}$, yielding $\widehat{\mathbf{F}}^{s}=D(\widehat{\mathbf{H}}^{s})\in\mathbb{R}^{T_s\times d}$, which is matched to $\mathbf{F}^{s}$ to implicitly guide the semantic compressor with the linguistic information in the ASR features. It is defined as
\begin{equation}
    \mathcal{L}_{\mathrm{FRec}}
    = \left\|\widehat{\mathbf{F}}^{s}-\mathbf{F}^{s}\right\|_2^2,
\label{eq:frec}
\end{equation}

\noindent\textbf{Codec losses.}
The codec losses consist of reconstruction, adversarial, and quantization terms. For reconstruction, a multi-scale mel-spectrogram loss $\mathcal{L}_{\mathrm{mel}}$ between $\mathbf{x}$ and $\widehat{\mathbf{x}}$ is used, following DAC~\cite{dac}. For adversarial training, the generator is optimized with a least-squares GAN (LSGAN)~\cite{lsgan} loss $\mathcal{L}_{\mathrm{adv}}$ and a feature-matching loss $\mathcal{L}_{\mathrm{fm}}$ against a multi-period discriminator (MPD)~\cite{hifigan} and a multi-resolution discriminator (MRD)~\cite{univnet}, while the discriminators are trained separately with the LSGAN objective. For quantization, RVQ is trained with a codebook loss $\mathcal{L}_{\mathrm{cb}}$ and a commitment loss $\mathcal{L}_{\mathrm{cm}}$~\cite{dac}. The codec loss is
\begin{equation}
\begin{aligned}
    \mathcal{L}_{\mathrm{codec}}
    ={}& \lambda_{\mathrm{mel}}\mathcal{L}_{\mathrm{mel}}
    + \lambda_{\mathrm{adv}}\mathcal{L}_{\mathrm{adv}}
    + \lambda_{\mathrm{fm}}\mathcal{L}_{\mathrm{fm}} \\
    &+ \lambda_{\mathrm{cb}}\mathcal{L}_{\mathrm{cb}}
    + \lambda_{\mathrm{cm}}\mathcal{L}_{\mathrm{cm}},
\end{aligned}
\label{eq:codec}
\end{equation}

\noindent\textbf{Overall objective.}
Combining these losses, the overall objective for Q-SPT is
\begin{equation}
    \mathcal{L}
    = \mathcal{L}_{\mathrm{codec}}
    + \lambda_{\mathrm{AR}}\mathcal{L}_{\mathrm{AR}}
    + \lambda_{\mathrm{QRec}}\mathcal{L}_{\mathrm{QRec}}
    + \lambda_{\mathrm{FRec}}\mathcal{L}_{\mathrm{FRec}}.
\label{eq:total_loss}
\end{equation}
We set $(\lambda_{\mathrm{mel}},\lambda_{\mathrm{adv}},\lambda_{\mathrm{fm}},\lambda_{\mathrm{cb}},\lambda_{\mathrm{cm}})=(15,1,2,1,0.25)$ following FlexiCodec~\cite{flexicodec}, and $(\lambda_{\mathrm{AR}},\lambda_{\mathrm{QRec}},\lambda_{\mathrm{FRec}})=(20,15,100)$. The text embedding, text FFN, LM head, and feature decoder $D$ are used only during training.

\begin{table*}[t]
\centering
\begingroup
\fontsize{9}{10}\selectfont
\renewcommand{\arraystretch}{1.0}
\setlength{\tabcolsep}{2pt}
\setlength{\aboverulesep}{0.3ex}
\setlength{\belowrulesep}{0.3ex}

\caption{Speech reconstruction results on LibriSpeech test-clean.
$^*$ denotes our stride-modified DualCodec.
Bold indicates the best result within each frame-rate group.}
\label{tab:main_recon}

\begin{tabular*}{\textwidth}{@{\extracolsep{\fill}}l*{11}{c}@{}}
\toprule

& &
\multicolumn{2}{c}{\textbf{Bitrate (kbps)}} &
\multicolumn{2}{c}{$\boldsymbol{n_q=1}$} &
\multicolumn{6}{c}{$\boldsymbol{n_q=8}$} \\
\cmidrule(lr){3-4}
\cmidrule(lr){5-6}
\cmidrule(lr){7-12}

\textbf{Model}
& \textbf{Hz}
& $n_q=1$
& $n_q=8$
& \textbf{WER}$\downarrow$
& \textbf{SIM}$\downarrow$
& \textbf{WER}$\downarrow$
& \textbf{PESQ-NB}$\uparrow$
& \textbf{PESQ-WB}$\uparrow$
& \textbf{MCD}$\downarrow$
& \textbf{SIM}$\uparrow$
& \textbf{UTMOS}$\uparrow$ \\
\midrule

Mimi~\cite{moshi}
& 12.5 & 0.138 & 1.100
& 76.78 & 0.09 & 2.96
& 2.90 & 2.25
& 4.28 & 0.79 & 3.34 \\

OmniCodec~\cite{omnicodec}
& 12.5 & 0.138 & 1.100
& 69.05 & \textbf{0.07} & 2.53
& 2.83 & 2.15
& 3.75 & 0.85 & 3.36 \\

LM-SPT~\cite{lmspt}
& 12.5 & 0.175 & 1.225
& \textbf{3.38} & 0.26 & \textbf{2.21}
& 3.41 & 2.76
& 3.12 & \textbf{0.91} & 3.84 \\

DualCodec~\cite{dualcodec}
& 12.5 & 0.175 & 1.225
& 5.19 & 0.48 & 2.25
& \textbf{3.42} & \textbf{2.82}
& \textbf{3.01} & 0.87 & \textbf{3.97} \\

\midrule

DualCodec$^*$
& 6.25 & 0.088 & 0.613
& 48.95 & 0.27 & 4.26
& 2.80 & 2.12
& 3.87 & 0.72 & 3.86 \\

FlexiCodec~\cite{flexicodec}
& 6.25 & 0.113 & 0.638
& 4.58 & 0.22 & 2.68
& 2.84 & 2.19
& 3.55 & 0.73 & 3.86 \\

\textbf{Q-SPT}
& 6.25 & 0.094 & 0.619
& \textbf{4.29} & \textbf{0.19} & \textbf{2.52}
& \textbf{2.90} & \textbf{2.26}
& \textbf{3.52} & \textbf{0.77} & \textbf{4.07} \\

\bottomrule
\end{tabular*}

\endgroup
\end{table*}
\begin{table}[t]
\centering
\begingroup
\fontsize{9}{10}\selectfont
\renewcommand{\arraystretch}{1.0}
\setlength{\tabcolsep}{2.5pt}
\setlength{\aboverulesep}{0.3ex}
\setlength{\belowrulesep}{0.3ex}

\caption{SLM results on LibriSpeech test-clean at 6.25\,Hz.
ASR uses the first code stream, and TTS uses all eight streams.
$^*$ denotes our stride-modified DualCodec.
Best results are shown in bold.}
\label{tab:main_slm}

\begin{tabular*}{\columnwidth}{@{\extracolsep{\fill}}lcccc@{}}
\toprule
& \multicolumn{1}{c}{\textbf{ASR}}
& \multicolumn{3}{c}{\textbf{TTS}} \\
\cmidrule(lr){2-2}
\cmidrule(lr){3-5}

\textbf{Model}
& \textbf{WER}$\downarrow$
& \textbf{WER}$\downarrow$
& \textbf{UTMOS}$\uparrow$
& \textbf{DNSMOS}$\uparrow$ \\
\midrule

DualCodec$^*$
& 120.16 & 14.72 & 3.07 & 3.18 \\

FlexiCodec
& 6.18 & \textbf{10.65} & 3.05 & 3.10 \\

\textbf{Q-SPT} (w/o text loss)
& 7.04 & 12.31 & 3.16 & 3.24 \\

\textbf{Q-SPT}
& \textbf{5.88} & 11.82 & \textbf{3.44} & \textbf{3.28} \\

\bottomrule
\end{tabular*}

\endgroup
\end{table}
\begin{table}[t]
\centering
\begingroup
\fontsize{9}{10.5}\selectfont
\renewcommand{\arraystretch}{1.0}
\setlength{\tabcolsep}{0.5pt}
\setlength{\aboverulesep}{0.3ex}
\setlength{\belowrulesep}{0.3ex}

\caption{Ablation study of Q-SPT on speech reconstruction.
Shared boundaries are applied only at inference.
Best results are shown in bold.}
\label{tab:ablation}

\begin{tabular*}{\columnwidth}{@{\extracolsep{\fill}}lccccc@{}}
\toprule
& &
\multicolumn{1}{c}{$\boldsymbol{n_q=1}$} &
\multicolumn{3}{c}{$\boldsymbol{n_q=8}$} \\
\cmidrule(lr){3-3}
\cmidrule(lr){4-6}

\textbf{Configuration}
& $\boldsymbol{\mathcal{L}_{\mathrm{AR}}}$
& \textbf{WER}$\downarrow$
& \textbf{WER}$\downarrow$
& \textbf{PESQ-WB}$\uparrow$
& \textbf{UTMOS}$\uparrow$ \\
\midrule

Shared weights
& \xmark & 6.16 & 2.86 & 2.16 & 4.02 \\

Shared boundaries
& \cmark & \textbf{4.29} & 3.83 & 2.14 & 3.99 \\

Separate compressors
& \xmark & 4.97 & 2.59 & 2.25 & 4.03 \\

Separate compressors
& \cmark & \textbf{4.29} & \textbf{2.52}
& \textbf{2.26} & \textbf{4.07} \\

\bottomrule
\end{tabular*}

\endgroup
\end{table}
\section{Experiments}
\label{sec:experiments}

\subsection{Experimental Setup}
\label{sec:training_setup}

\noindent\textbf{Datasets.} For codec training, we use speech--text pairs with utterance durations of at most 30\,s from LibriHeavy~\cite{libriheavy}, GigaSpeech~\cite{gigaspeech}, and the English subset of Emilia~\cite{emilia}, totaling about 47.7k hours. All speech is resampled to 16\,kHz. SLMs are trained on the full 960-hour LibriSpeech training set~\cite{librispeech}. All experiments are evaluated on LibriSpeech test-clean.

\noindent\textbf{Implementation details.}
Q-SPT has 104.1M parameters at inference, excluding the frozen ASR encoder. Each compressor consists of three Transformer layers. The semantic FSQ has a vocabulary of 32,768 codes, and each of the seven acoustic RVQ codebooks has 4,096 entries. The codec is trained for 950k steps on 8 NVIDIA L40S GPUs using AdamW with a learning rate of $10^{-4}$ and dynamic batches of up to 30\,s of audio per GPU.

For SLM experiments, we fully fine-tune a separate Qwen3.5-4B~\cite{qwen35} for each codec, jointly on ASR and TTS in a 1:1 ratio, using tokens extracted from the frozen codec. ASR predicts text from the first stream, whereas TTS generates all eight streams in parallel with a delay pattern~\cite{musicgen} that shifts each stream by one frame. Each SLM is trained for 20k steps with a global batch size of 128 using AdamW with a learning rate of $2\times10^{-4}$. FlexiCodec-based TTS additionally predicts per-token frame lengths with a duration head~\cite{flexicodec}.

\noindent\textbf{Baselines.}
At 12.5\,Hz, we compare with Mimi~\cite{moshi}, OmniCodec~\cite{omnicodec}, LM-SPT~\cite{lmspt}, and DualCodec~\cite{dualcodec}. At 6.25\,Hz, we compare with FlexiCodec~\cite{flexicodec} and DualCodec. We adapt DualCodec to 6.25\,Hz by modifying its stride settings and train it on the same data as Q-SPT. For FlexiCodec, we use the official threshold $\tau=0.867$, which yields a nominal frame rate of 6.25\,Hz.

\noindent\textbf{Evaluation metrics.}
We denote the number of code streams used for reconstruction by $n_q$, where $n_q=1$ uses only the semantic stream and $n_q=8$ uses all eight streams. We report word error rate (WER) and speaker similarity (SIM) at $n_q=1$ and $n_q=8$, and additionally report Perceptual Evaluation of Speech Quality in narrowband and wideband modes (PESQ-NB/WB)~\cite{pesq}, mel-cepstral distortion (MCD)~\cite{mcd}, and UTMOS~\cite{utmos} at $n_q=8$. For SLMs, we report ASR and TTS WER, together with UTMOS and DNSMOS~\cite{dnsmos} for TTS quality. Reconstruction and TTS WER are computed with a HuBERT-Large ASR model~\cite{hubert}, and SIM with a fine-tuned WavLM-Large speaker-verification model with an ECAPA-TDNN backend~\cite{wavlm,ecapa_tdnn}. At $n_q=1$, lower SIM indicates less speaker information retained in the semantic stream, whereas higher SIM is preferred at $n_q=8$. Bitrates are computed from the frame rate and codebook sizes, including three bits per output frame for FlexiCodec's frame-length information~\cite{flexicodec}.

\subsection{Main Results}
\label{sec:main_results}

\noindent\textbf{Codec reconstruction.}
Table~\ref{tab:main_recon} compares the reconstruction performance of speech codecs. At 6.25\,Hz, Q-SPT achieves the best results on all metrics, including the lowest WER at both $n_q=1$ and $n_q=8$. DualCodec performs strongly at 12.5\,Hz, but when adapted to 6.25\,Hz, its WER at $n_q=1$ increases from 5.19 to 48.95 and its acoustic quality also degrades. This suggests that average-pooling-based compression struggles to preserve information at extremely low frame rates. FlexiCodec preserves linguistic information better, with a WER of 4.58 at $n_q=1$, but shows lower quality than Q-SPT on all $n_q=8$ quality metrics, even at a slightly higher bitrate. This is consistent with its semantic-derived merging boundaries being applied to the acoustic stream. Q-SPT also yields the lowest SIM at $n_q=1$ among the 6.25-Hz codecs and higher SIM than FlexiCodec at $n_q=8$, indicating that speaker information is effectively separated from the semantic stream and recovered through the acoustic residual streams. Notably, at roughly half the bitrate, Q-SPT achieves higher UTMOS than all 12.5-Hz codecs.

\noindent\textbf{Downstream speech modeling.}
Table~\ref{tab:main_slm} shows the ASR and TTS performance of SLMs trained on the tokens of each codec. For ASR, which uses only the semantic tokens, Q-SPT outperforms all baselines with a WER of 5.88, indicating that it preserves linguistic information that the SLM can effectively exploit. In contrast, DualCodec$^{*}$ fails to produce meaningful transcriptions, suggesting that the limitation of average-pooling-based compression carries over to downstream modeling. For TTS, Q-SPT achieves the best perceptual quality in both UTMOS and DNSMOS, improving UTMOS from 3.05 to 3.44 over FlexiCodec. FlexiCodec, which additionally uses a duration head, achieves a lower WER but lower perceptual quality. This trend is consistent with Table~\ref{tab:main_recon}, where FlexiCodec also shows lower full-stream perceptual quality than Q-SPT.

\subsection{Ablation Study}
\label{sec:ablation}
To analyze how each design choice contributes to Q-SPT, we conduct three ablations in Table~\ref{tab:ablation}. First, we share the compressor parameters across the two streams to test whether each stream benefits from its own parameters. Second, we match the acoustic window boundaries to those of the semantic compressor, analogous to the shared merging boundaries in FlexiCodec, to examine how reconstruction quality changes. Third, we remove the autoregressive text loss to measure the effect of explicit linguistic supervision. The first and third variants are trained with the same setup as Q-SPT, whereas the second is applied only at inference.

\noindent\textbf{Shared weights.}
We share the Transformer layers of the two compressors while keeping the temporal windows of each stream. The shared-weight variant uses six layers to match the parameter count of the two three-layer compressors, and both variants are trained without the autoregressive text loss. Sharing the layers degrades both streams, increasing WER at $n_q=1$ from 4.97 to 6.16 and worsening WER and PESQ-WB at $n_q=8$, while UTMOS remains nearly unchanged. This indicates that separate compressor parameters are more effective than shared parameters for modeling the semantic and acoustic streams.

\noindent\textbf{Shared boundaries.}
To examine how sharing temporal boundaries across streams affects reconstruction, we match the acoustic window boundaries to those of the semantic compressor. At inference, each acoustic window is shifted to end at the same point in time as the corresponding semantic window, while the window size and all trained parameters remain unchanged. As the semantic stream is unchanged, WER at $n_q=1$ remains at 4.29, whereas at $n_q=8$, WER increases from 2.52 to 3.83 and PESQ-WB and UTMOS also decrease. This indicates that the trained acoustic compressor relies on its own temporal boundaries, and that imposing the semantic boundaries at inference degrades acoustic reconstruction.

\noindent\textbf{Text supervision.}
To measure the effect of explicit linguistic supervision, we compare separate-compressor models trained with and without $\mathcal{L}_{\mathrm{AR}}$. Adding $\mathcal{L}_{\mathrm{AR}}$ reduces WER at $n_q=1$ from 4.97 to 4.29, while slightly improving all $n_q=8$ metrics. The benefit also transfers to downstream SLMs, reducing ASR WER from 7.04 to 5.88 and TTS WER from 12.31 to 11.82 (Table~\ref{tab:main_slm}). Without $\mathcal{L}_{\mathrm{AR}}$, Q-SPT already outperforms FlexiCodec on all reported \(n_q=8\) reconstruction metrics, but not on WER at $n_q=1$ (4.97 vs.\ 4.58).
Thus, autoregressive text supervision complements the proposed query-based compression by further strengthening linguistic information in the semantic tokens.

\section{Conclusion}
We proposed Q-SPT, a low-frame-rate dual-stream speech tokenizer that replaces rule-based temporal compression with learnable query-based compression and supervises the semantic compressor with an autoregressive text loss. At 6.25\,Hz, Q-SPT achieves the best reconstruction among the evaluated codecs, while effectively separating speaker information from the semantic stream. In addition, the autoregressive text loss further strengthens linguistic information in the semantic tokens. These benefits carry over to downstream SLMs, where Q-SPT outperforms the baselines in ASR accuracy and TTS perceptual quality while maintaining competitive intelligibility.

\clearpage
\begingroup
\ninept
\bibliographystyle{IEEEbib}
\section{Acknowledgments}
This work was supported by the Institute of Information \& Communications Technology Planning \& Evaluation (IITP) grant funded by the Korea government (MSIT) (No. RS-2019-II190079, Artificial Intelligence Graduate School Program (Korea University)).
\bibliography{refs}
\endgroup

\end{document}